\documentstyle[aps,preprint,tighten,floats,epsf,rotate]{revtex}

\def\be{\begin{equation}}
\def\ee{\end{equation}}

\begin{document}


\preprint{\vbox{\it 
                        \null\hfill\rm    SI-TH-99-1, hep-ph/9901296}\\\\}
%

\title{Formation of extended topological defects during symmetry 
breaking phase transitions in $O(2)$ and $O(3)$ models}
\author{G. Holzwarth\thanks{%
e-mail: holzwarth@physik.uni-siegen.de}}
\address{Fachbereich Physik, Universit\"{a}t Siegen, 
D-57068 Siegen, Germany} 
\date{January 1999}
\maketitle
\begin{abstract}
The density of extended topological defects created during symmetry 
breaking phase transitions depends on the ratio between the 
correlation length in the symmetric phase near $T_c$ and 
the winding length of the defects as determined by the 
momentaneous effective action after a typical relaxation time.
Conservation of winding number in numerical simulations 
requires a suitable embedding of the field variables and the 
appropriate geometrical implementation of the winding density 
on the discrete lattice. We define a modified Kibble limit for the 
square lattice and obtain defect densities as functions of 
winding lengths in $O(2)$ and $O(3)$ models. The latter allows to 
observe formation of disoriented aligned domains within the easy
plane. Their extent is severely limited by the momentaneous defect
density during the course of the quench.
\end{abstract}
\pacs{PACS numbers: }

\draft

\vspace{1 cm}
\section{Introduction}

Formation of defects during symmetry breaking phase transitions has
found increasing attention in a variety of applications ranging from
condensed matter systems to cosmological scenarios~\cite{General}. 
The idea to consider baryons as topological defects in an $N$-component
chiral meson field $\Phi_i$ constrained by $\Phi^2 =$ const.~\cite{SkyWi},
similarly has initiated attempts to estimate multiplicities of
baryon and antibaryon production in high energy events or heavy ion
collisions from the dynamics of defect formation during the cooling 
phase of an expanding hot hadronic gas~\cite{Ellis}\cite{KaSri}. 
There are strong theoretical
indications that during this cooling process the chiral $O(N)$ symmetry 
is spontaneously broken near a critical temperature $T_c$, and
since the critical temperature is estimated to be of the order of the
pion mass $m_\pi$ (\cite{GeLeu} and refs. therein), 
it may appear sufficient to describe this chiral 
transition in the framework of low-energy effective theory for the
mesonic fields $\Phi_i$~\cite{ChPT}. However, perturbative methods for the
evaluation of the relevant effective potential near the critical
temperature are probably not very reliable, the role eventually
played by other degrees of freedom is not really understood, 
so at present it is not
even clear whether this chiral transition would be first or 
second order (see e.g. the discussion in~\cite{PiWi}\cite{KoKo}). 

In analogy to phenomena observed in condensed matter systems there also
have been speculations about the simultaneous ocurrence of extended
domains with different average orientation of the aligned field 
('disoriented chiral condensate' (DCC)) which
rearrange into a uniform vacuum on a much longer time scale and
therefore could manifest themselves through anomalous branching ratios
for the production of differently charged mesons~\cite{Anselm}. 
Evidently, the existence of differently oriented domains and defect
formation are intimately related, especially if the defects are of
topological type which cannot simply disappear from a field
configuration by local unwinding.

Depending on the physical system under consideration 
the nature of the defects is determined 
by the choice of the manifold
on which the fields live~\cite{Polyakov}: 
If the constraint $\Phi^2=$ const. is enforced,
the fields live on the $(N-1)$-dimensional sphere $S^{N-1}$,
conveniently parametrized by angular variables. 
This embedding can provide
defects with a topologically protected winding number and thus 
in case of the chiral field allows to identify them with baryons.

Near the phase transition, however, it may be appropriate to relax the
constraint and allow all $N$ components of $\Phi$ to move independently.
Then a Euclidean $R^N$ embedding appears most convenient and in fact,
the possible formation of DCC's in the chiral phase transition
has been investigated in this framework~\cite{Raja}. 
The winding number of a field configuration embedded in $R^N$
is, however, not topologically conserved: it
will undergo discrete changes, if at some space-time point all field
components vanish, i.e. if the field configuration moves across the
origin $\Phi^2(x)=0$. 

In order to preserve the identification of winding number with 
baryon number it is necessary to insist on the angular nature
of the chiral field, 
while the additional degree of freedom is picked up by the modulus
$|\Phi|$, the scalar $\sigma$-field. 
In other words, the topologically trivial $R^N$ embedding is
replaced by the $R^1\times S^{N-1}$ manifold with nontrivial homotopy
group $Z$. Here the origin $\Phi=0$ is excluded as a highly singular 
branch point where
different angular sheets are tied together. Any field configuration
which moves across this point, leaves a defect or antidefect of winding
number $\pm 1$ at the spatial position where this happens, which remains
connected to the rest of the moving field by a string. The spatial
structure of string and defect are determined by the dynamics of the
classical field. The nontrivial structure of the modulus of $\Phi$
along the string constitutes the 'bag', which interpolates from the
vacuum value of the $\sigma$-field surrounding the defect to
a (small) value in its center where the angular fields change rapidly
from one sheet to the next. For a simple 1+1 dimensional $O(2)$ model
this has been discussed in detail in~\cite{gh}. 

These considerations naturally pose two questions: How do 
different embeddings affect the average multiplicities of defects and
antidefects and the possible existence of disoriented domains during
the chiral transition? And: What is the influence of the spatial
structure of the topological defects (as determined by the
temperature-dependent effective action)
on their production multiplicities and
on the possible formation of large extended disoriented domains? 

One may hope that for the essential features of defect formation the
order of the transition is not very important.
If the transition is first order then it would proceed through nucleation
of bubbles~\cite{KaVi}, which in their interior are characterized by an aligned
chiral field with nonvanishing average value 
$ \langle |\Phi| \rangle = f_0(T_c)$ of 
the chiral field as determined through the nontrivial minimum of the 
effective potential near $T_c$.
The orientation of the aligned field in different bubbles could be
considered as random, so the average distance of their centers can be
taken as an initial correlation length $\xi(T_c)$ at $T_c$. This length
provides the relevant scale for defect formation, irrespective of its 
physical size. On this scale the total volume $V$ considered can be 
identified with the total number of bubble seeds near $T_c$. 
Growth and coalescence of
bubbles then leads to the formation of topological defects within
a typical formation time $\tau$ characteristic for the damping of local
fluctuations. 

Decisive for the multiplicity of the created defects is the
ratio between their spatial extent, i.e. the 'radius' or 'winding length'
$l_W$ of the defects at the time of their formation
and the correlation length $\xi(T_c)$.
The winding length is determined by the stabilization
mechanism supplied through the effective action for the average
field.  

If the winding length $l_W$ is much smaller than the
correlation length the defects can be
considered as pointlike on the scale $\xi(T_c)$.
Then the purely combinatorial rules~\cite{Kib}
of random lattices apply for their formation, i.e. their average
density $n$ is given by the Kibble limit 
\be
\label{kibble}
n=(N_++N_-)/V=\frac{1}{2}\;(2)^{-D}\;\lambda_D 
\ee
where $(N_++N_-)$ is the number of defects plus antidefects,
$D$ is the space dimension, and the factor $\lambda_D$ is the average
ratio of the numbers of $D$-simplices and vertices in a large random 
lattice, i.e. $\lambda_1=1, \lambda_2=2,
\lambda_3=\frac{24}{35}\pi^2,$ etc.~\cite{Christ}. 
In this case the random field
fluctuations during the typical time $\tau$ get transformed into a 
rather dense ensemble of defects and 
antidefects. These may become diluted in the following due to annihilation 
processes on a much larger timescale.

If, however, the value of the winding length $l_W$ 
is larger than the average distance 
of the bubble seeds, then stabilization dynamics will prevent the 
formation of pointlike defects and antidefects inside
the volume occupied by large extended stable defects or antidefects and
(\ref{kibble}) has to be replaced by
\be
\label{modkib}
n=(N_++N_-)/V
=\frac{1}{2}\;(2\frac{l_W}{\xi})^{-D}\;\lambda_D.
\ee
i.e. before they have a chance to form, point defects 
and antidefects get instantly (i.e. within time $\tau$) eaten by big
stable solitons which carry the net winding number of the chiral
field within the volume occupied by them. The formation of these
large extended solitons is a direct process which does not proceed
through multiple annihilations of smaller defects. 

If for a sudden quench ($t_Q \ll \tau$)
the effective potential at $T=0$ is established
before any dynamical or dissipative mechanism can affect the initial random
field configuration then the winding length $l_W(T)$ near $T=0$ will be 
relevant for the size of the 'cold' defects formed. With the relevant
scale still given by the initial correlation length or 
average bubble distance $\xi(T_c)$ near $T_c$, the ratio
$l_W(T=0)/\xi(T_c)$ will enter into (\ref{modkib}).  

If typical quench times are much larger than the
formation time ($t_Q \gg \tau$) then the number of initially created
defects and antidefects will be given by (\ref{modkib}) with 
winding length $l_W$ characteristic for $T \approx T_c$.
Subsequent changes in the defect multiplicities during the 
continuation of the quench depend on the scaling behaviour of their
winding length: if $l_W(T)$ decreases with the decreasing temperature
(as in three spatial dimensions where $l_W$ scales like $f_0(T)^{-1}$) 
then the (initially small) number of 
(initially large) defects stays small, so the multiplicity remains near
the ratio $l_W(T_c)/\xi(T_c)$ in (\ref{modkib}).
On the other hand, if $l_W(T)$ increases during the quench 
(as in one spatial dimension where $l_W$ scales like $f_0(T)^{1/2}$), 
the growing defects swallow many of the defect-antidefect
pairs, so at the end of the slow quench the resulting multiplicities
are the same as for a sudden quench, i.e. determined by the ratio
$l_W(T=0)/\xi(T_c)$. 

If the transition is second order no bubbles appear and the
average distance between bubbles should be
replaced by an instantaneous correlation length $\xi(T)$. For
finite rate of cooling $\xi(T)$ remains finite near $T_c$ because 
critical slowing down prevents response over large
distances~\cite{Zurek}. 
On the other hand $f_0(T)$ stays close to zero which (in 3 dimensions)
implies large $l_W$.
So also in this case the relevant winding lengths may exceed the
pertinent correlation length of the plasma, so defect multiplicities
can be much smaller than the limiting value given by eq. (\ref{kibble}).

Defect formation is intimately related with the size of DCC domains.
If the transition proceeds in a way as to form a dense gas of pointlike
defects and antidefects according to the Kibble limit (\ref{kibble}), then 
domains of topologically trivial mesonic field which fill the
remaining space between the different defects
consequently have to be small (in units of $\xi$). 
With increasing ratio $l_W/\xi$ the
number of defects created decreases rapidly according to
(\ref{modkib}), but the chiral field in the spatial 
domains separating the defects is correlated due to the large winding
length of the defects which 'organize ' the field surrounding them.
In both limiting cases there seems to be little room for large 
(in units of $\xi$) domains of
aligned but randomly oriented locally trivial field configurations. 
The actual physical size of these domains, of course, still depends on the
magnitude of the basic length scale $\xi$ in physical units; 
this, naturally, depends on the physical system considered
and on the order of the transition. For the hot chiral gas in the
chirally symmetric phase above $T_c$ numbers for $\xi$ typically 
quoted range from 0.5 fm to 1 fm. On a lattice with a lattice constant
of that size individual 'cold' baryons will cover just one lattice
unit or less, while on formation near $T_c$ the 'melted' baryons~\cite{Hans} 
may smoothly 
extend over a large number of cells. We therefore expect a situation 
where the above considerations about the relevance of the winding
length typically apply.

For a discussion of the chiral phase transition in the framework of
effective meson fields it is therefore necessary to embed the fields in
a manifold which allows for the topological protection of winding
number, i.e. to use a separation into angular field variables and one
modulus $\sigma$-field. 
Practically, in lattice simulations, where topological arguments no
longer apply, conserving algorithms have to be used for the updating
of configurations during their evolution. We will describe the results
of such simulations for the 1d-$O(2)$ model in section II, for the
2d-$O(3)$ model in section III.  It is expected that
the bias introduced into the initial configurations through explicit
chiral symmetry breaking provides another efficient mechanism to 
suppress defect formation~\cite{Dziarmaga}. This effect should be discussed in
connection with the evolution of the $\sigma$ field.
We shall concentrate here on the interplay between
winding length and defect densities. Because the winding
length is mainly determined by the dynamics of the angular field
variables we postpone a detailed discussion of features related
to the evolution of the modulus $\sigma$-field and, correspondingly,
use unbiased initial configurations only.

\section{$O(2)$-model in 1+1 dimensions}

Some of these features can be nicely visualized in the simple $O(2)$
model in 1+1 dimensions. As discussed above we concentrate here on the
relation between winding length and defect density. To define the
notation and for completeness we repeat some facts discussed
in~\cite{gh} concerning the 
topologically trivial $R^2$ vs. nontrivial $R^1 \times S^1$ embedding.
Defects arise as stable static solutions for the effective lagrangian density
taken in the standard $\Phi^4$ form (summation over $a$=1,2 and
$\mu$=0,1 understood)
\be
\label{lagrange}
{\cal L}_0(\Phi) = \frac{1}{2} \partial_\mu \Phi_a  \partial^\mu \Phi_a
+ \frac{\lambda_c}{4} (\Phi^2 - f^2)^2 - H \Phi_1 \; .
\ee
The term $H \Phi_1$ in (\ref{lagrange}) breaks chiral
$O(2)$ symmetry explicitly.
In order to keep the minimum of the potential for finite $H$ at 
the $H$-independent value $\Phi^2 = f^2_0$ we define
\be
f^2 = f^2_0 - \frac{H}{\lambda_c f_0} \;.
\ee
In accordance with conclusions from 1-loop renormalization group 
applied to (\ref{lagrange}) we will take $\lambda_c$ and $H$ as temperature
independent, while $f_0(T)$ decreases as $T$ rises from $T=0$
towards $T=T_c$. 

The cartesian components $\Phi_{1,2}$ ($-\infty<\Phi_{1,2}<\infty$)
define the Euclidean $R^2$ embedding of this model, while modulus field
$\sigma$ and the angular variable $\phi$
($0<\sigma<\infty, -\infty<\phi<\infty$) related to $\Phi_i$ through
\be
\label{sigpi}
\Phi_1(x,t)=\sigma(x,t)\;\cos\phi(x,t),~~~~~~~~
\Phi_2(x,t)=\sigma(x,t)\;\sin\phi(x,t)
\ee 
define the embedding into the $S^1\times R^1$ manifold.
The $\sigma$- and $\pi$-masses corresponding to these fields are
\be
\label{masses}
m_\pi^2=\frac{H}{f_0},~~~~~~~m_\sigma^2=2 \lambda_c f_0^2 + m_\pi^2.
\ee
A convenient unit for dimensionful quantities is the
temperature-independent symmetry breaker $H$ of mass dimension 2.
So, in the following, if numbers are given for dimensionful quantities,
they are to be multiplied by appropriate powers of $H$.

In the Euclidean $R^2$ embedding the static part of
(\ref{lagrange}) leads to 
stable nontopological solitons as long as the inequality
\be
\label{ineq}
 f_0^3 \frac{\lambda_c}{H} > 21.28 
\ee
is satisfied. Otherwise, the only stable solution is the trivial ground
state $\Phi_1\equiv f_0, \; \Phi_2\equiv 0$.
This corresponds to approximately $m_\sigma > 6.6\; m_\pi$, as
stability condition.
For a typical coupling strength $\lambda_c/H=60$ (which we use in the
following) this implies $m_\pi<1.2$ .
On the other hand, in $R^1 \times S^1$,
the winding number $B$ of topologically nontrivial field configurations
with fixed angular boundary conditions 
$\phi(\pm \infty,t)=2 \pi\: n_\pm$,~~($n_\pm:$ integer)
\be
\label{bnum}
B=\frac{1}{2\pi}\int_{-\infty}^{\infty} \frac{\partial}{\partial x}
\phi(x,t) dx = n_+-n_-,
\ee
is topologically conserved.
For values of $f_0(T)$ below the limiting value (\ref{ineq})
the angular field $\phi$ of nontrivial static configurations collapses into 
pointlike defects at some position $x=x_0$, while the modulus field
$\sigma$ describes a spatially extended 'bag' profile satisfying
\be
\label{radeq}
\sigma''-\lambda_c \sigma (\sigma^2-f^2)+H=0
\ee
$$
\mbox{with~~~~ } \sigma(x\to x_0) \to +0, \mbox{~~~~and~~~~}
\sigma(|x|\to \infty) \to f_0. 
$$
The radius of this bag scales approximately like $\pi/m_\sigma$. 
Therefore, with $f_0$ small near $T_c$, where the stability condition
(\ref{ineq}) is not satisfied and the angular field collapsed to a point
defect, the bag radius still is comparable
to the angular winding length $l_W\approx\pi/m_\pi$ of the 
soliton, which solves the static $O(2)$ model (\ref{lagrange}),
if the constraint $\Phi^2 \equiv f_0^2$ is enforced, i.e., if we simply
consider a U(1) model on a circle $S^1$. 

On the other hand, if the stability condition (\ref{ineq}) is satisfied,
the static solutions of the $O(2)$ model in $R^2$ and in $R^1 \times S^1$
embedding coincide. Because the modulus field of these
configurations stays close to $f_0$, the corresponding angular field 
essentially also coincides with the soliton of the 
model constrained to the circle $S^1$ with $\sigma\equiv f_0$.

Statistical ensembles of these stable solutions are expected at 
the end of a cooling process that evolves from an initial ensemble of random
configurations with correlation length $\xi$. The quench is defined
through the variation of $f_0(T)$ with time. 
For a numerical simulation the initial configurations before the onset 
of the quench we choose Gaussian deviates in the
field components $\Phi_{1,2}(i)$, at the points
$x_i=i\;\xi,~~(i=0,..,N_L)$ of a spatial grid of total length $L=N_L \xi$
with lattice constant $\xi$ which defines the correlation
length or distance of bubble seeds near $T_c$. 
Periodic boundary conditions are imposed by choosing
$\Phi_{1,2}(0)=\Phi_{1,2}(N_L)$.  From these random values
of the cartesian field components the initial configuration of the
angular field $\phi(i)$ is obtained by taking the shortest path from
the angle at point $x_i$ to the angle at a neighbouring point $x_{i+1}$.
This guarantees that the absolute value of the increment in the angle from
one point to the next is always less than $\pi$. Due to the periodic
boundary conditions in the cartesian field components the difference in
the angular field $\phi(N_L)-\phi(0)$ then is an integer multiple of $2\pi$,
which defines the winding number $B$ of that particular random initial
configuration. 

As we discussed above the winding length of nontrivial solutions in
different embeddings is similar to the winding length of the model
with the modulus of $\Phi$ constrained to $\Phi^2 \equiv f_0^2(T)$.
The essentials of the relation between defect winding length and defect
density therefore can also be observed within the simple U(1) (or
nonlinear $O(2)$) model.
As an example we consider the sudden quench where the initially
prepared configurations are exposed at time $t=0$ to the effective action with
constant $f_0(T=0)$. Then, within relaxation time $\tau$, the
configurations evolve into an ensemble of stable defects with winding length
$l_W\approx\pi/m_\pi=\pi (f_0/H)^{1/2}$.
In fig.1 the average final density $n$ of defects plus
antidefects averaged over an ensemble of 100 random initial configurations
selected with total winding $B=0$ on an $N_L$=1000 lattice
is plotted against the symmetry-breaking mass $m_\pi$.

\begin{figure}[h]
\begin{center}
\leavevmode
\vbox{\epsfxsize=10truecm \epsfbox{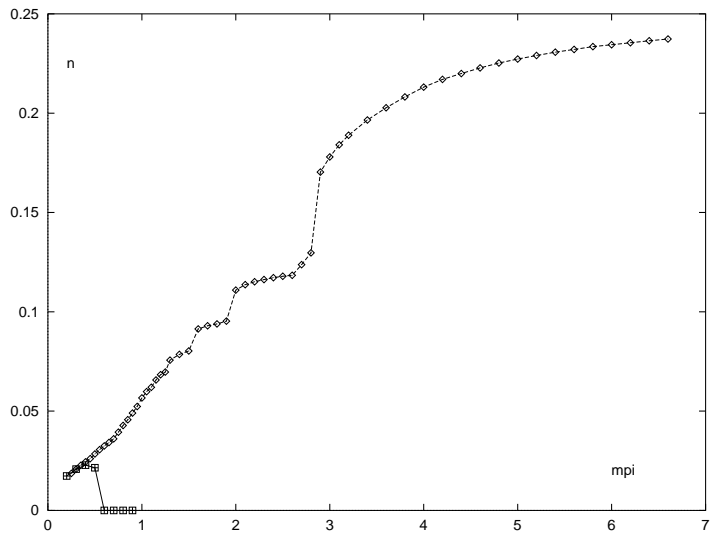}}
\end{center}
\caption{ Defect plus antidefect density $n=(N_++N_-)/N_L$ averaged over
an ensemble of 100 random initial configurations on an $N_L=1000$ lattice
after a sudden quench as function of inverse winding length $m_\pi$ for
the nonlinear $O(2)$ model. The squares in the lower left corner show
the same observable for the linear $O(2)$ model in $R^2$-embedding. } 
\label{fig1} 
\end{figure}

The region where $m_\pi$ is less than $\pi$ is essentially
characterized by a linear
rise of the density $n$ as expected from (\ref{modkib}) for $D=1$. 
Beyond this value the winding length becomes
comparable to the lattice constant $\xi$ and we observe the onset of
saturation with $n$ slowly converging towards the Kibble limit
(\ref{kibble}) of $n=0.25$ (for $D=1$).  
A peculiar feature is that the increase in the number of finally
surviving defects and antidefects with decreasing winding length
proceeds in quite well-defined plateaus which are very robust against
details of the evolution algorithm. Note that the absolute number of
defects in the ensemble underlying fig.(1) is rather large: e.g. in
the sudden decrease of $n$ just below $m_\pi=2.9$
a total average number of $0.05 \cdot 10^5$ defects are eaten by a tiny
increase of the winding length, while in the region ($2.1\le m_\pi\le
2.8$) mainly empty space in between defects is reduced. Evidently, this
is an indication of a lattice effect: as we have noted, if
$m_\pi$ drops below $\pi$ the defects start to grow beyond one lattice
unit. As expected, for very small values of $m_\pi$ the average density
is again rather smooth. Altogether, we may conclude that the experiment
confirms the expectation expressed in (\ref{modkib}).

The squares connected by a full line in the left lower corner of fig.1 show
average defect densities as calculated in the unconstrained
$O(2)$ model in Euclidean $R^2$ manifold. 
The theoretical limit of defect stability here (for
$\lambda_c=60$) is close to $m_\pi \approx 1.2 $. However, in the finite
lattice calculation the defects disappear already near  $m_\pi \approx 0.6$.
Beyond that value the resulting defect
densities are zero, while in the region of stability the average
densities closely follow the linear rise as obtained in the $U(1)$ model. 
Irrespective of the stability condition, for the unconstrained $O(2)$ 
model embedded in $R^1\times S^1$, the defect
densities are similar to those of the $U(1)$ model for all values of
$m_\pi$. However, in
the region of instability where (\ref{ineq}) is violated, average
multiplicities in $R^1 \times S^1$ embedding slightly exceed the $U(1)$
results due to the smaller size of the radial bag (depending on the
choice of $\lambda_c$) (cf. eq. (\ref{masses})).

Relations between winding length $l_W$ of defects and formation of DCC domains
cannot be addressed within the $O(2)$ model, because the strength of
the symmetry breaker necessary for finite $l_W$ prevents local
alignment in random directions. 
Outside the range of the defects the angle $\phi$ stabilizes always
at multiples of $2\pi$. We therefore turn to the $O(3)$ model whose
additional freedom allows to at least discuss partial aspects of 
DCC formation.

\section{Nonlinear $O(3)$-model in 2+1 dimensions}

Corresponding to our discussion of the constrained $O(2)$ model we
consider in this section only that version of the $O(3)$ model where
the modulus of $\Phi$ is constrained to a fixed value $f_0$. In two
spatial dimensions $f_0$ is of mass dimension 1/2, so we use appropriate
powers of $f_0^2$ as units for all dimensionful quantities.
Then the fields $\Phi$ are unit vectors $\hat{\Phi}$ which
live on the sphere $S^2$ and conveniently are parametrized by two angles
$\theta$ and $\phi$
\begin{eqnarray}
\label{thetphi}
\Phi_1(\mbox{\bf{x}},t)&=&
\cos\phi(\mbox{\bf{x}},t)\sin\theta(\mbox{\bf{x}},t) \nonumber \\
\Phi_2(\mbox{\bf{x}},t)&=&
\sin\phi(\mbox{\bf{x}},t)\sin\theta(\mbox{\bf{x}},t) \nonumber \\
\Phi_3(\mbox{\bf{x}},t)&=&
\cos\theta(\mbox{\bf{x}},t)
\end{eqnarray}
If $\hat{\Phi}$ converges towards a fixed unit vector, say  $\hat{e_3}$,
for spatial infinity on $R^2$
(sufficiently fast for the energy to converge) then the configuration
space for smooth fields is disconnected due to the nontriviality
of the second homotopy group $\pi_2(S^2)=Z$. 
Individual configurations then are characterized by the winding number
$B$ and local winding density $\rho$
\be
\label{B}
B=\int \rho \; d^2x ,
\ee
\be
\label{rho}
\rho\equiv\frac{1}{8 \pi}\epsilon_{ij}\hat{\Phi}\cdot(\partial_i 
\hat{\Phi} \times \partial_j \hat{\Phi}) 
\ee
(summation over spatial indices $i,j$ understood).

We consider the effective energy functional 
\be
\label{baby}
{\cal E}[\hat{\Phi}] 
=\int \;  \left( \frac{1}{2} \partial_i \hat{\Phi} \cdot  
\partial_i \hat{\Phi}
+\lambda^2 \;\rho^2+ \frac{1}{2 \lambda^2} 
\; (\hat{\Phi}-\hat{e_3})^2 \right) d^2x .
\ee
Evidently, $\lambda$ serves to scale the spatial extension of stable
solutions, as it can be eliminated by $x \to \lambda x$.  
Both terms in ${\cal E}$, the
$\rho^2$ ('Skyrme') term and the explicitly symmetry-breaking last term, are
necessary to stabilize solitons (the 'baby skyrmions'~\cite{Piette}) 
with fixed winding length $l_W \propto \lambda$. For our present purpose to
study the relation between $l_W$ and average defect densities created
in a symmetry breaking transition it is therefore sufficient to 
impose the quench through an appropriate time dependence of $\lambda$.

Discretizing
the spatial coordinates to a two-dimensional lattice, the limit
$\lambda \ll 1$ will produce defects which are pointlike on the scale of the
lattice constant, while $\lambda \gg 1$ will create smooth density
distributions which extend over many lattice units.
A characteristic problem arises in numerical simulations on a 
discrete lattice if the spatial extent
of solitons is of the order of a few lattice constants or even less:
this implies, that differences in the angles $\theta$ and $\phi$ between
neighbouring lattice points may be of the order of $\pi$. In that case
the definition (\ref{rho}) for the winding density $\rho$ no longer applies.
It is only for infinitesimally small differentials that 
the surface element $d\Omega$ on $S^2$ which is the image of the elementary
lattice cell $d^2x$ is given by $d\Omega = 4 \pi \rho \; d^2x$. But it is
essential to maintain also for finite lattice constants 
the geometrical meaning of $d\Omega$ as the surface area 
which the image of the unit lattice cell cuts out from $S^2$. 
This can be readily
implemented by defining $2\pi\rho$ as the (oriented) area of the spherical
triangle which is cut out on $S^2$ by the three geodesics which connect
the end points of three $\hat\Phi$-vectors attached to three 
corners of an elementary lattice cell. 
This works for arbitrary relative orientations of these $\hat\Phi$-vectors
and allows even to detect one complete soliton inside one elementary
lattice cell\footnote{In other words, we use a triangulation of the plane in
terms of the triangles formed by two adjoining lattice links and one diagonal 
in each square lattice cell. Depending on the choice of the diagonal in
the unit cell the area covered by the image of that
cell on $S^2$ may differ by $\pm 4 \pi$, i.e. the winding number may
differ by $\pm 1$. However, for the ensemble average, this is not
relevant.}. Clearly, this definition reduces to (\ref{rho}) in the
continuum limit. Periodic boundary conditions (which compactify $R^2$
to a torus), or $\hat\Phi \to \hat e_3$ (which compactifies $R^2$ to
$S^2$) then lead to integer values of the total winding number $B$
obtained by summing up all oriented spherical triangles.

The topological considerations which guarantee conservation of
$B$ for continuous fields do not hold on the
discrete lattice, because $\hat\Phi$-vectors attached to neighbouring
points on the lattice can differ arbitrarily. However, $B$-conservation
can easily be re-implemented into the evolution of the configurations
by allowing in each updating step only configurations\footnote{
Practically, in a local updating procedure this requires to
control the sum of $\rho$ only in the local vicinity of the points which
are affected by the update.} which conserve
$B$. In fact, this is a convenient way to
compare evolutions which conserve $B$ with others that allow for local
unwinding of defects. So it is also not necessary to introduce specific
types of potentials (see e.g.~\cite{Ward}) into (\ref{baby}) 
to avoid the 'exceptional'
configurations~\cite{Luesch} as doorways for unwinding.

For the first term in (\ref{baby}) there is no similar compulsory
extension to finite lattice constants because it is not of geometrical
nature. So, as usual we just interprete it as nearest neighbour interaction.

Having thus defined the implementation of the model (\ref{baby}) on a 
discrete lattice which conserves those features essential for our 
present purpose, it remains to specify the initial configurations
before the onset of the quench. Again we postpone here the question of
any possible bias which may exist in these initial sets due to explicit
symmetry breaking. Instead we choose a uniformly random set of 
$\hat\Phi$-vectors on all interior lattice points. Only on the
boundaries we impose $\hat\Phi =\hat e_3$, i.e. $\theta = 0$ with
random $\phi$.

In the following sense this initial random set realizes the maximal 
average number of defects which can be accommodated on this square
lattice: First it should be noted that by definition of our map
(which maps a triangle formed by two adjoining orthogonal 
lattice links and the diagonal connecting their endpoints 
onto the interior of the {\it smaller} of the
two complementary spherical triangles formed by the corresponding 
geodesics on $S^2$) the image triangle covers always {\it less} than half of
the sphere, and thus by definition does not contain a defect. So, this
is different from the Kibble mechanism which allows for defects inside
elementary triangles by considering an additional lattice point inside
the triangle. However, if we add the (oriented) areas covered by the 
two adjoining images of both triangles which make up the elementary 
square lattice cell then it may happen that this sum covers more than
half of the sphere, and by definition we then may count that as one defect
inside this elementary lattice cell. In fact, if $\lambda$ is very 
small, such a configuration will rapidly evolve into a configuration
where the sum of these two spherical triangles (which represent the
image of the lattice cell) is close to $4\pi$. We may call this the
Kibble limit for the square lattice. It differs from the triangular
limit (\ref{kibble}) by a factor $1/4$ (in two dimensions), which we
interprete as a difference in the definition of the correlation length 
$\xi$ by a factor
of two for uniformly random configurations on the vertices of a 
square lattice, as compared to the original Kibble counting which uses
one additional point with random field asignment inside each
of the two triangles which constitute the elementary square lattice
cell. We consider the identification of $\xi$ with the lattice constant
for a random asignment of field vectors to each vertex of the square
lattice as the natural definition of the correlation length.

This basic setup opens a wide variety of dynamical situations 
for systematic investigation. As just one example we shall discuss
here only the case of a linear quench in the scale parameter $\lambda$
\be 
\label{quench}
\lambda=\lambda_i+\frac{\lambda_f-\lambda_i}{t_Q} \; t~~~~~~~~~~~
(0\le t \le t_Q)
\ee
with $\lambda_i\gg 1$ and $\lambda_f \ll 1$. 
Starting from the initial random set, at the beginning of the
quench a large value of $\lambda_i$ attempts to establish long range
order with very small values of the local winding density $\rho$.
When $\lambda$ approaches values aound $\lambda\approx 3$ to $1$, 
individual defects start to be formed out of this more or less
organized low-density soup. Near the end of such a
quench, for very small $\lambda$ 
we expext an ensemble of $N_+$ defects and $N_-$ antidefects
each essentially localized on a single lattice cell. For $t > t_Q$ the
evolution can still be continued with $\lambda=\lambda_f$ to allow
for final stabilization of the defects formed, or for possible mutual
annihilations on a much longer time scale.

After the onset of the quench ordering of the configurations
proceeds within typical relaxation times $\tau$ and the extent to which
the ordering can spread before the local defects appear is governed by
the ratio $\tau/t_Q$, which is the appropriate measure for the quench
velocity. For a fast quench we expect the final defect density to be
close to our modified Kibble limit discussed above, because most of the
initial random (incomplete) winding will directly evolve into complete
local $4\pi$ winding. For slow quenches the random winding will have
time to be smoothed away into large areas of low winding density, before
finally a few defects and antidefects reappear. 

This naturally provides
the appropriate setting to investigate formation and size of DCC domains. 
There is, however, a severe limitation to this discussion, also within the
model (\ref{baby}): The stabilization mechanism of the baby skyrmions
relies on a strong symmetry breaker, the last term in (\ref{baby}). 
This term prevents formation of large areas of aligned but randomly
oriented field configurations if we consider the complete field vectors
$\hat\Phi$. In the spatial domains outside the range of the topological
defects the field will essentially be aligned in $\hat e_3$ direction.
Still, a partial aspect of DCC formation can be studied: In the so
called 'easy plane' orthogonal to $\hat e_3$, the field components 
$\Phi_1$ and $\Phi_2$ can freely align in random directions and the
partial correlation function
\be
\label{corr}
C(|2-1|)=<\cos\phi(1)\cos\phi(2)+\sin\phi(1)\sin\phi(2)>
\ee
(where the arguments 1,2 stand for two arbitrary lattice points)
provides the information about the size of aligned domains. It should,
however, be noted that the aligning force in the easy plane due to 
the first term in (\ref{baby}) (the nearest neighbour interaction) 
is active only in areas where $\sin \theta \not\equiv 0$. This means
that the formation of DCC domains can only be studied during the early
parts of the quench where the strength of the symmetry breaker is not
yet sufficient to form small localized defects with strictly aligned
(in $\hat e_3$ direction) domains in between. As soon as these local
defects appear (around $\lambda < 3$) the disoriented aligned areas in
the easy plane begin to re-randomize due to thermal fluctuations. 
So, the formation of DCC domains is closely connected to small but 
non-vanishing winding density, which by itself is an interesting
aspect. 
\begin{figure}[h]
\begin{center}
\leavevmode
\hbox{\epsfysize=6truecm \epsfbox{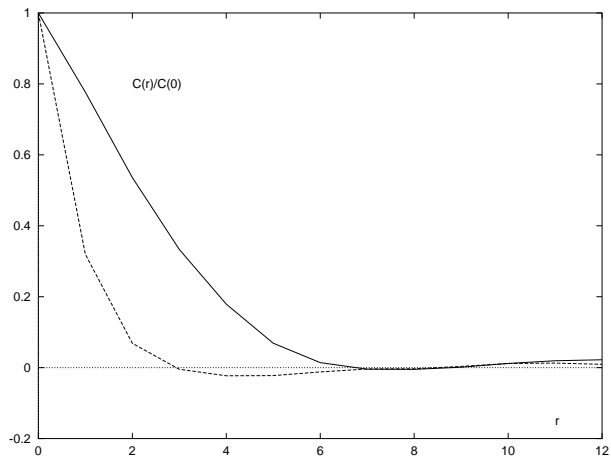}}
\end{center}
\caption{Correlation function (\ref{corr}) $C(r)/C(0)$ 
for fast ($t_Q=10$) (dashed line) and slow ($t_Q=200$) (full line) 
quench near $\lambda \approx 3$.}
\label{fig2} 
\end{figure}
\begin{figure}[h]
\begin{center}
\leavevmode
\hbox{\epsfysize=6truecm \epsfbox{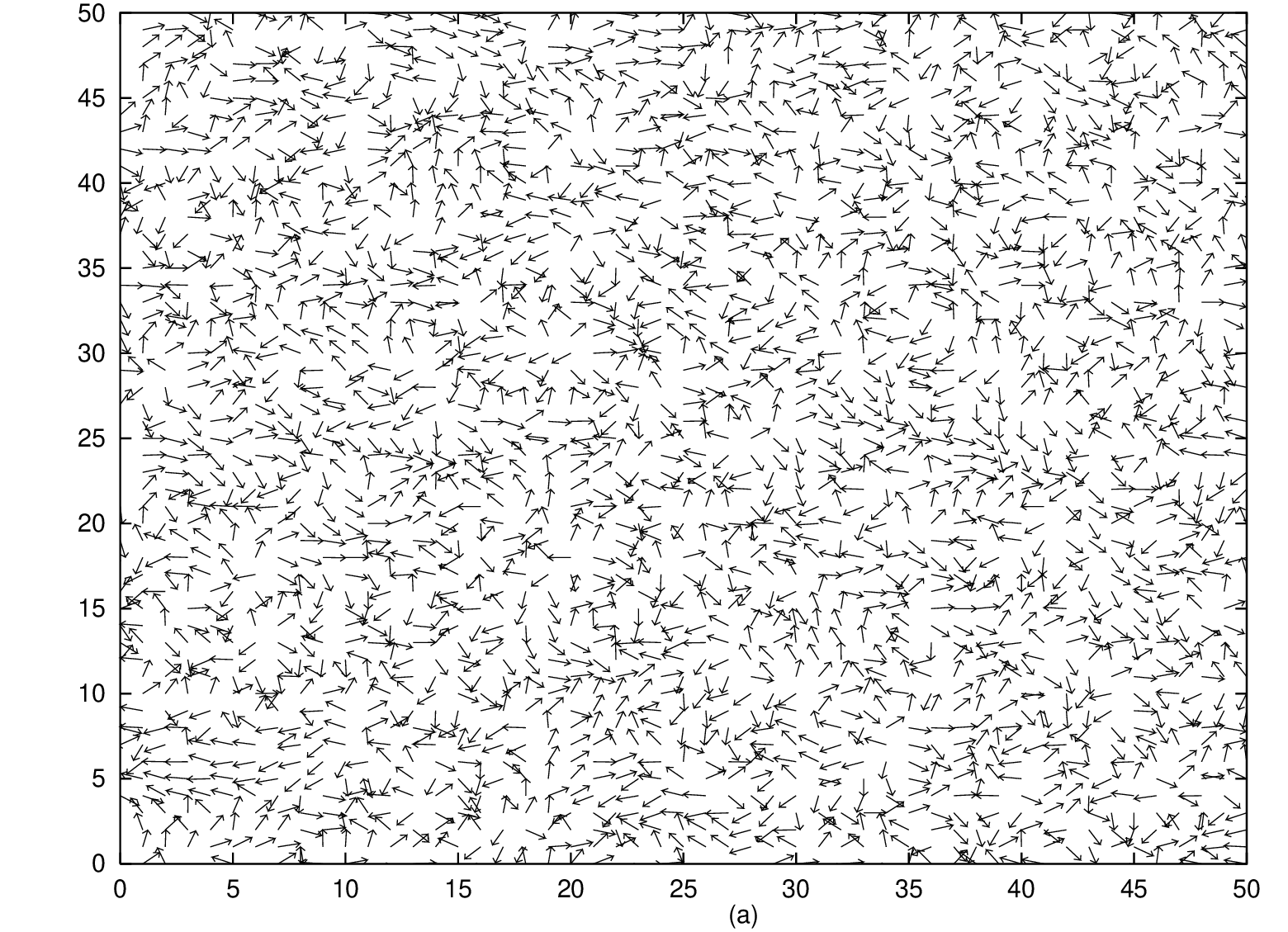} 
\epsfysize=6truecm \epsfbox{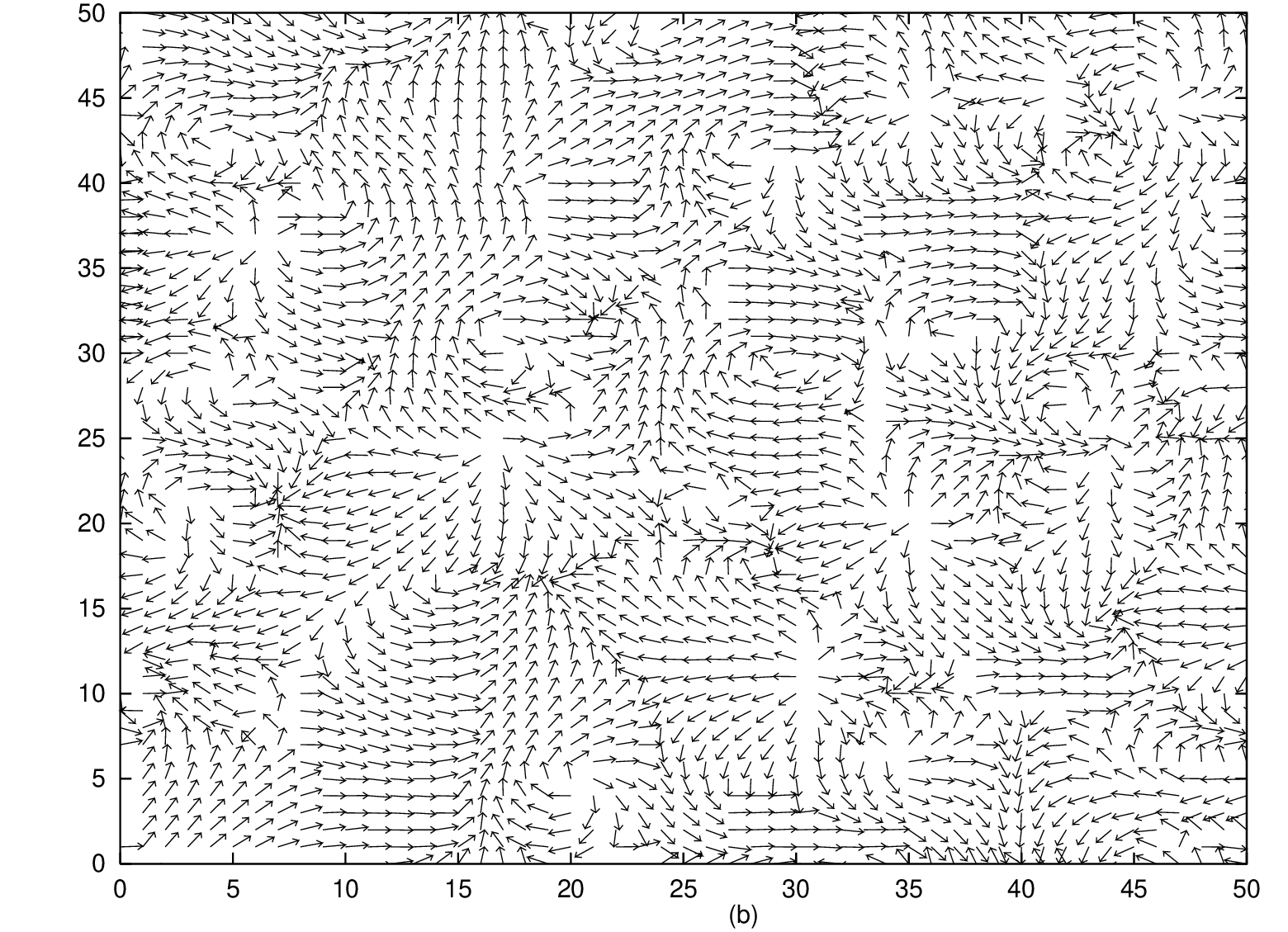}}
\end{center}
\caption{$\Phi$-field projected into (and re-normalized in) the easy plane 
for (a) fast ($t_Q=10$) 
and (b) slow ($t_Q=200$) quench near $\lambda \approx 3$.}
\label{fig3} 
\end{figure}
\begin{figure}[h]
\begin{center}
\leavevmode
\hbox{\epsfysize=6truecm \epsfbox{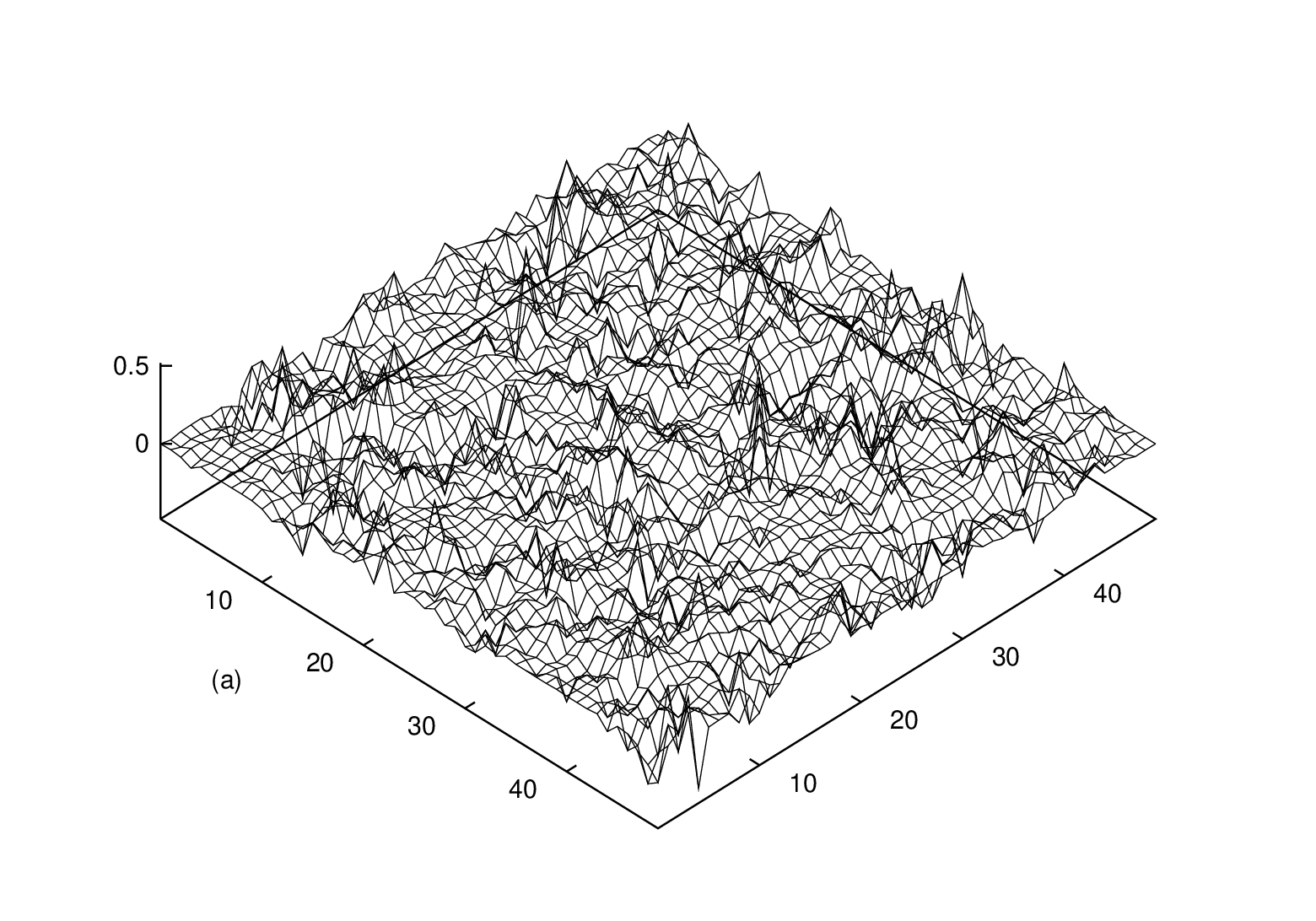} 
\epsfysize=6truecm \epsfbox{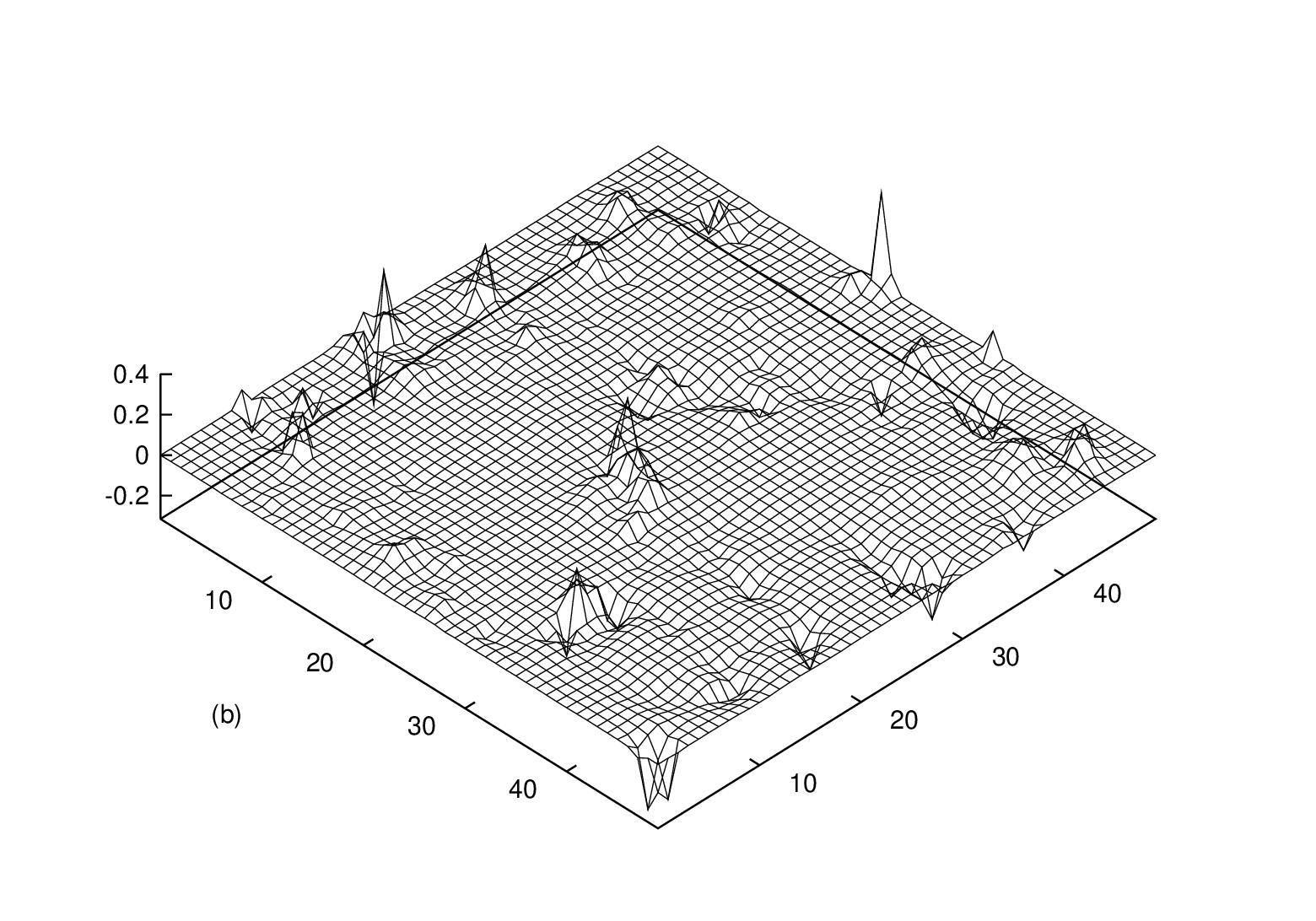}}
\end{center}
\caption{Winding density $\rho$   
during (a) fast ($t_Q=10$) and (b) slow ($t_Q=200$) quench
near $\lambda \approx 3$.}
\label{fig4} 
\end{figure}
\begin{figure}[h]
\begin{center}
\leavevmode
\hbox{\epsfysize=6truecm \epsfbox{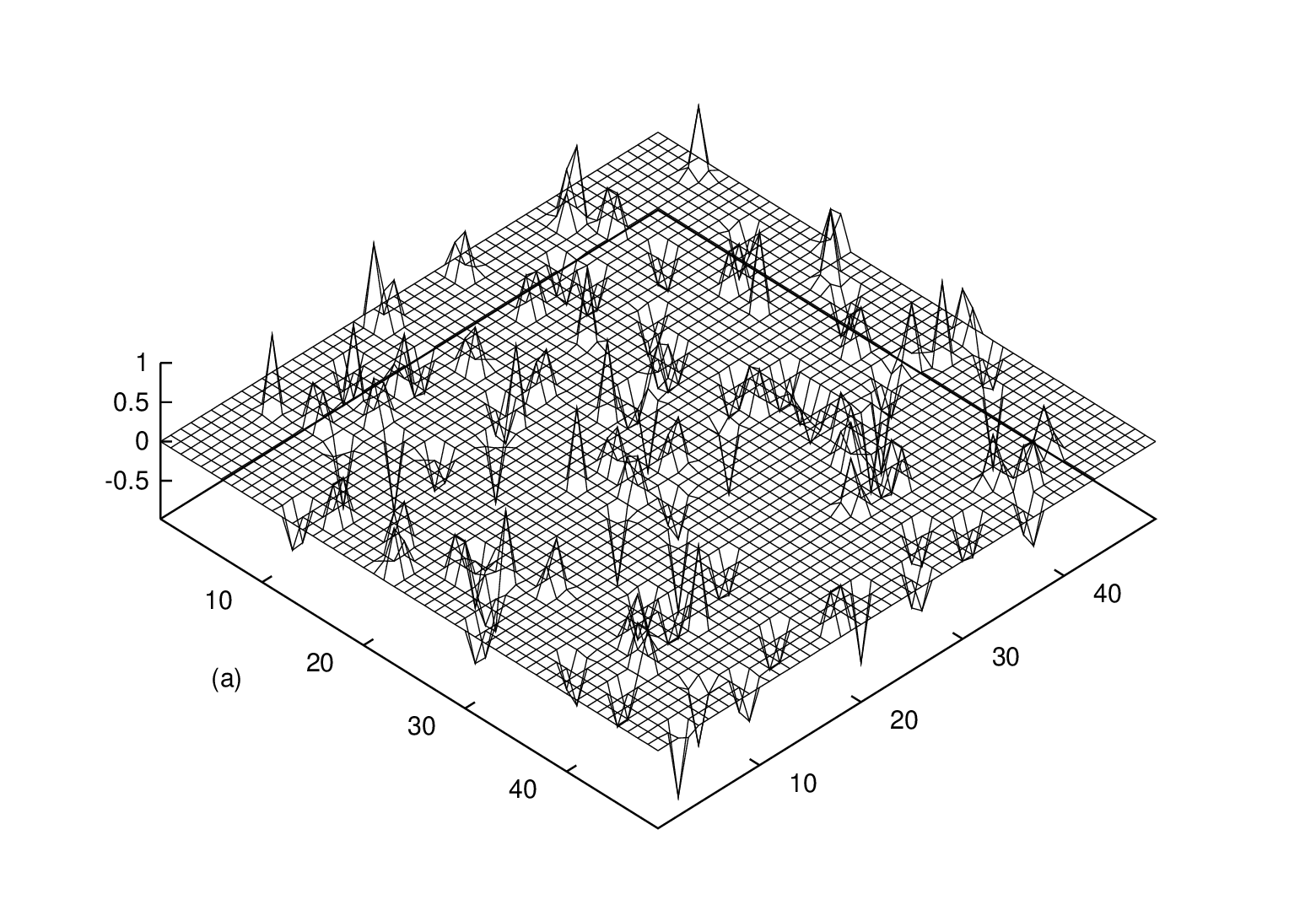} 
\epsfysize=6truecm \epsfbox{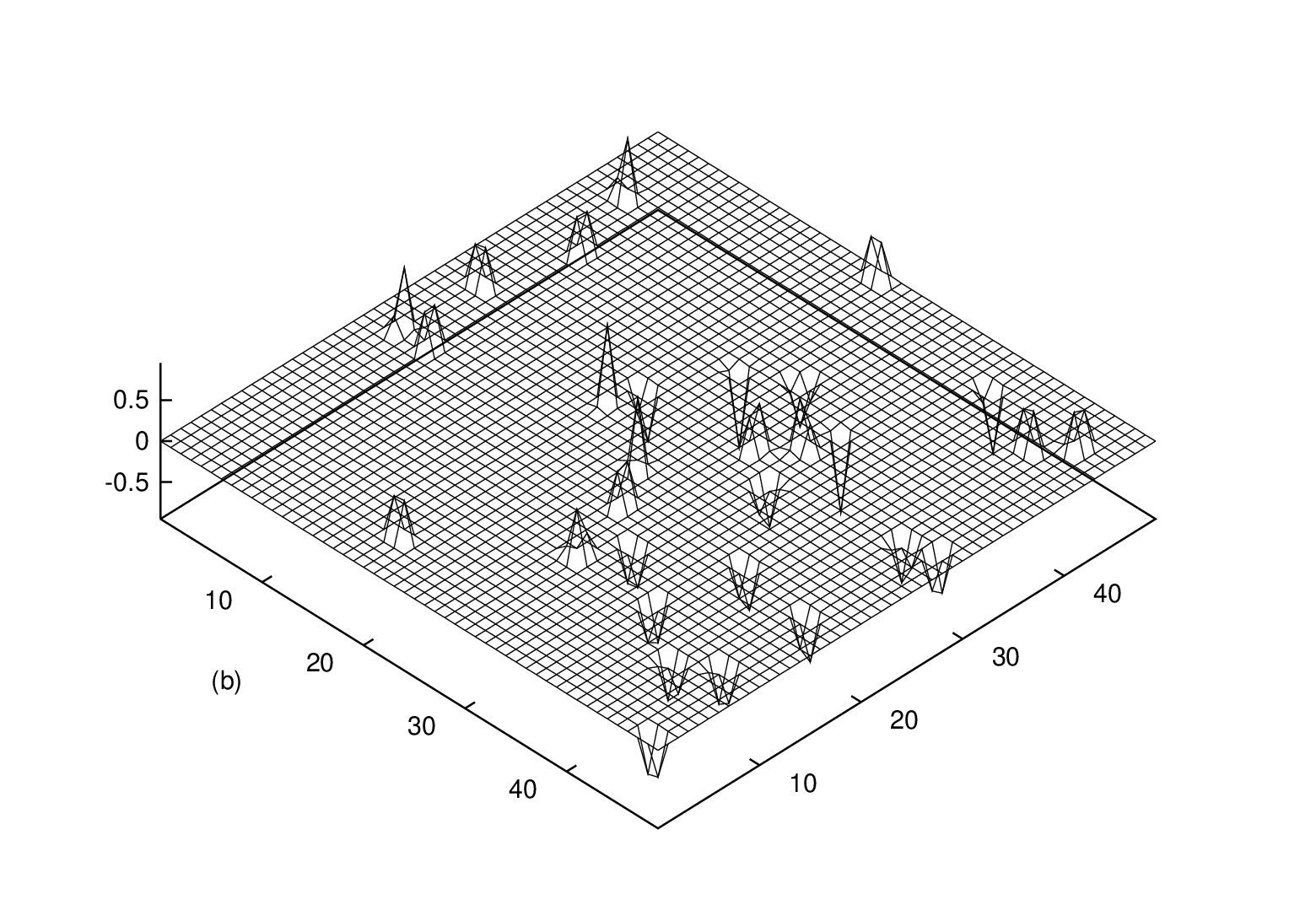}}
\end{center}
\caption{Final winding density $\rho$   
after (a) fast ($t_Q=10$) and (b) slow ($t_Q=200$) quench
at $\lambda=0.1$. The resulting (anti)defect numbers are 
$N_+=N_-=46$ and $N_+=N_-=15$ in cases (a) and (b), respectively.}
\label{fig5} 
\end{figure}
\begin{figure}[h]
\begin{center}
\leavevmode
\hbox{\epsfysize=12truecm \epsfbox{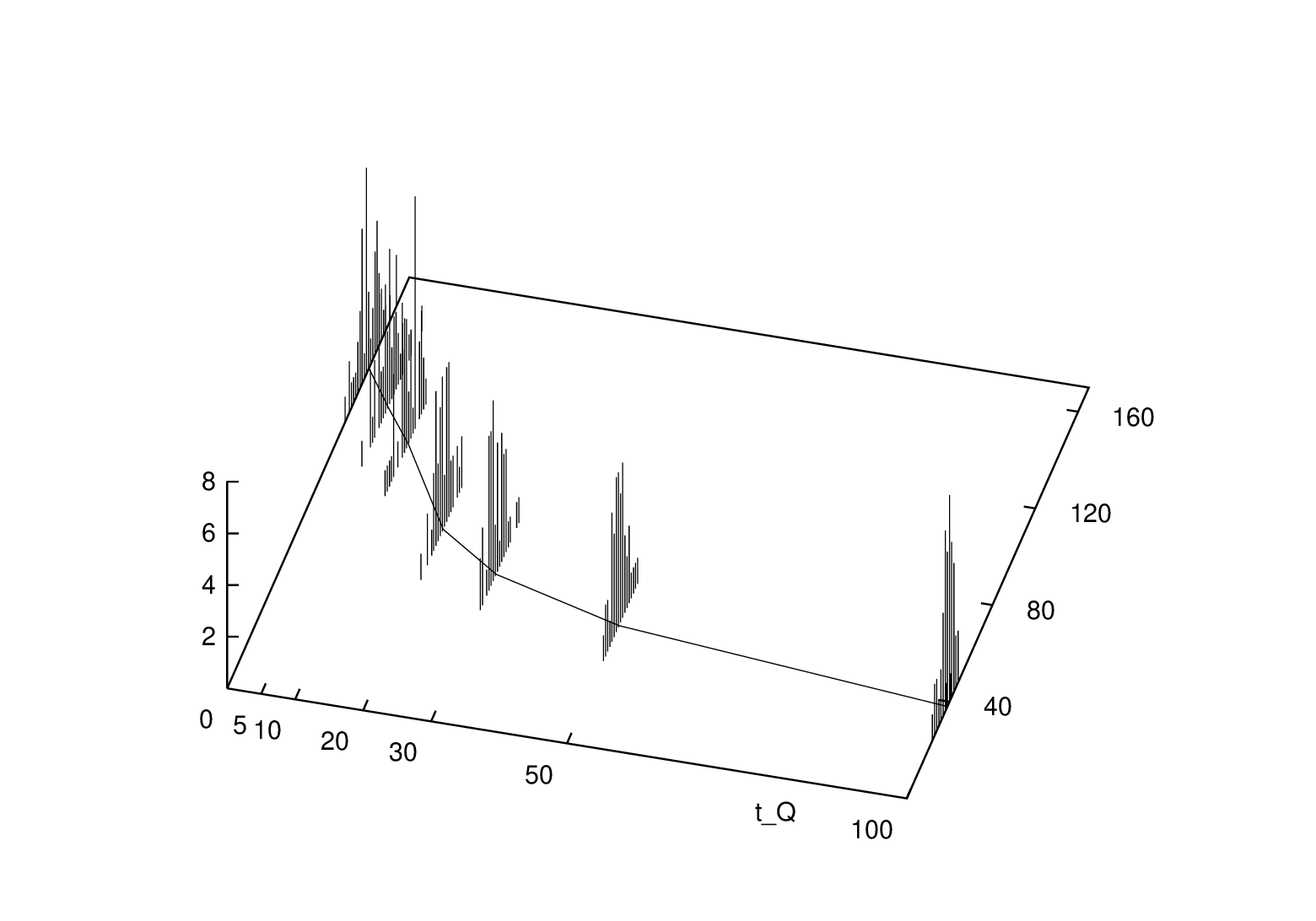}}
\end{center}
\caption{Final defect plus antidefect multiplicities 
for different quench times $t_Q$; their average values
are connected by the full line.}
\label{fig6} 
\end{figure}

In the following we describe some results of numerical
simulations on a 51$\times$51 lattice with an ensemble of random 
initial configurations which, however, have been selected as to satisfy
$B=0$. Of course, other values of $B$ can be selected to study a possible
influence of the total winding number. The algorithm conserves $B$ as
discussed above.
For illustration we present two intermediate steps in the evolution
of one specific arbitrary configuration during 
a fast quench ($t_Q = 10$ timesteps) and a slow quench
($t_Q = 200$ timesteps). 

 The first set shows the situation shortly before
$\lambda(t)$ crosses the value of $\lambda\approx 3$, i.e.
just before localized defects begin to form.  
The corresponding correlation functions (\ref{corr}) in
fig.2 indicate that the ordering in the easy plane has hardly
progressed beyond one lattice unit during the fast quench, while it
extends beyond five lattice units after the slow quench. This becomes
very evident if we look (in figs.3) at the 
field $\Phi$ projected into the easy
plane and normalized to unity in the easy plane (i.e. the components of
the arrows in the plot simply are $\cos \phi$ and $\sin \phi$). 
This provides an instant view of DCC domains and defects and it shows
that for the fast quench a high density of nascent defects leaves no 
space for large 
aligned domains while during the slow quench a well-developed
DCC-pattern emerges. The corresponding winding density
distributions $\rho$ display a rough surface (fig.4a) versus
essentially smooth
areas of low density with few emerging local spikes (fig.4b).
Finally, figs.5 show the resulting
ensemble of defects shortly after the end of the quench where
$\lambda$ has reached its final value of $\lambda_f=0.1$ and the 
local defects have stabilized.
Counting the final number of defects and antidefects which evolve from
this particular initial configuration
we find $N_+=N_-=46$ for the fast ($t_Q=10$) quench and $N_+=N_-=15$ 
for the slow ($t_Q=200$) quench.

Evolving $\cal N$ = 50 randomly chosen (but selected for
$B=0$) initial configurations through the quench (\ref{quench})
leads to an ensemble of defects and antidefects for each quench
time $t_Q$ as shown in fig.6. The modified Kibble limit as discussed
above for the square lattice is $N_++N_- \to (50\times50)/16 \approx
156 $. The boundary conditions ($\theta=0$) 
reduce this value to about 144 for this
lattice size. For the sudden quench ($t_Q=0$) the measured numbers approach
a mean value of only about 133 due to a few instant annihilations. 
For increasing
quench time the defect numbers drop to a mean value of about 32
near $t_Q=200$ as indicated by the full line in fig.6 which connects
the mean values for the different quench times.
The root mean square deviation from the average values decreases from about 
$\Delta \approx 10$ for the sudden quench to $\Delta \approx 5$ near
$t_Q=200$. Of course, these details depend on the specific choice
of the quench mechanism used here. For applications to specific
physical situations the appropriate temperature dependence of the
effective action and temporal structure of the quench has to implemented.

\section{Conclusion}
 
The purpose of this note is to study the influence of the
spatial extent of topological defects on their formation 
during a dynamical quench in a symmetry breaking phase transition.
Apart from the combinatorial factors of the Kibble mechanism it is the
ratio of the momentaneous magnitude of the winding length $l_W$ 
(as determined by the temperature dependent effective classical lagrangian) 
to the correlation length $\xi$ near
$T_c$ which is essential for the resulting multiplicities of
defects and antidefects. This leads to a crucial role of the quench
time on the scale of the relaxation time $\tau$ typically needed for
defect formation. If during the early part of the quench the 
winding length is large on the scale of $\xi$, the resulting defect
multiplicities are much reduced as compared to the combinatorial
Kibble limit for pointlike defects.

As illustration we presented numerical simulations in the nonlinear
1d-$O(2)$ and 2d-$O(3)$ models which are characterized by a 
topologically conserved winding number $B$, in view of applications
where this number corresponds to a physically relevant observable.
For the lattice simulation it is essential to implement the 
geometrical meaning of the winding density for the discrete map.
Then boundary conditions on angular
variables still serve to enforce integer values for $B$, although
topological arguments for $B$-conservation no longer apply. However,
by explicit control during the evolution $B$-violating steps can be
excluded.  
If conservation of $B$ is an important feature of the physical
system under discussion, like baryon number conservation in chiral
meson field models, extensions to the linear versions of $O(N)$
models have to include the additional degree of freedom in the form of
a modulus variable in an $S^{N-1}\times R^1$ embedding. 
For comparison we have briefly discussed
the topologically trivial $R^2$-embedding of the linear $O(2)$ model.
Unfortunately, in these low-dimensional models, the winding
length of defects is closely tied to explicit symmetry breaking.
This leads to a dramatic difference between Euclidean and angular embedding
if the symmetry breaking is sufficiently strong to destabilize
nontrivial configurations. 
The same feature also hampers the investigation of DCC domains in these
low-dimensional models. It is only in the easy plane of the $O(3)$
model where partial aspects of randomly oriented alignment can be
observed. 
It has been argued that the
bias introduced into the ensemble of initial configurations (for $ T>T_c$) by 
explicit symmetry breaking causes another drastic reduction of defect
multiplicities. This interesting aspect has been omitted here, again
for the same reason that the symmetry breakers simultaneously have to
serve as stabilizers for the defects.

Of course, the real challenge is the formation of 3d-$O(4)$ skyrmions 
as a nonperturbative model of how baryons emerge in the cooling of the 
hot meson soup. In that model stabilization of defects and explicit
symmetry breaking through small mesonic masses are almost unrelated.
Therefore all the interesting features concerning winding length, 
formation of DCC domains, effects of the
modulus field, baryon number conservation, bias in the initial 
ensemble, all can be studied quite independently with the same methods
used here for baby skyrmions.

\acknowledgements
The author would like to thank 
J. Klomfass, J. Dziarmaga and B.A. Ivanov for interesting discussions.

\end{document}